\documentclass[letterpaper,journal]{IEEEtran}
\usepackage{amsmath,amsfonts}
\usepackage{algorithmic}
\usepackage{algorithm}
\usepackage{array}
\usepackage[caption=false,font=normalsize,labelfont=sf,textfont=sf]{subfig}
\usepackage{textcomp}
\usepackage{stfloats}
\usepackage{url}
\usepackage{verbatim}
\usepackage{graphicx}
\usepackage{cite}
\usepackage[all]{nowidow}
\usepackage[dvipsnames]{xcolor}

\setnowidow[3]
\setnoclub[3]

\newcommand{\idro}{{\sf Idrogen}}
\newcommand{\WR}{{White Rabbit}}

\newcommand{\mtca}{{$\mu$TCA}}

\begin{document}

\title{Laser Synchronisation Over One Hundred Kilometers With Stability at Picosecond Scale}

\author{
    \IEEEauthorblockN{
        Alice Renaux\IEEEauthorrefmark{1},
        Ronic Chiche\IEEEauthorrefmark{1},
        Aurélien Martens\IEEEauthorrefmark{1},
        Antoine Back\IEEEauthorrefmark{1}, 
        Paul-Eric Pottie\IEEEauthorrefmark{2}, 
        and Daniel Charlet\IEEEauthorrefmark{1}
    }
    \\
    \IEEEauthorblockA{\IEEEauthorrefmark{1} 
        Laboratoire de Physique des 2 Infinis Irène Joliot-Curie (IJCLab), Université Paris-Saclay, CNRS/IN2P3\\
        Orsay, France 
    }
    \\
    \IEEEauthorblockA{\IEEEauthorrefmark{2} 
        Laboratoire Temps Espace (LNE-OP), Observatoire de Paris, Université PSL, Sorbonne Université, Université de Lille, LNE, CNRS\\
        Paris, France \\
        Email: aurelien.martens@ijclab.in2p3.fr        
    }
}

\markboth{Journal of \LaTeX\ Class Files,~Vol.~14, No.~8, August~2021}%
{Shell \MakeLowercase{\textit{et al.}}: A Sample Article Using IEEEtran.cls for IEEE Journals}

\IEEEpubid{0000--0000/00\$00.00~\copyright~2021 IEEE}

\maketitle

\begin{abstract}
Large-scale systems, such as very large accelerators used for fundamental research, require the implementation of precise timing and synchronization systems over distances of several kilometers. Femtosecond synchronisation has been reached by the implementation of costly and complex clock distribution systems. However, many devices, such as accelerator diagnostics or detectors for physics at colliders, only require picosecond stability and, in some cases, similar accuracy. An approach that is based on the CERN White Rabbit protocol, deployed on an electronic system capable of generating arbitrary frequencies with Hertz precision, is proposed here. Results of performance tests for the synchronization of a laser system, typically employed as a diagnostic for electron/positron beam polarimetry in accelerators, are provided in this Paper. We demonstrate that the system can synchronize a pulsed laser with picosecond stability over one hundred kilometers on the short-term. The long-term stability over half a day is found to be of 5.5~ps for the 100~km link. The accuracy of the phase difference corresponding to $\pm 20$~ps is obtained. This work paves the way for the deployment of White-Rabbit-based synchronization systems for accelerator components, such as lasers, but also for large-scale detectors.
\end{abstract}


\section{Introduction}
\IEEEPARstart{L}{arge} accelerators, as colliders, generally employed for fundamental research in particle physics, are of several kilometers in circumference. Existing example is the Large Hadron Collider (LHC) at CERN~\cite{LHC} with an approximate circumference of 27~km. The SuperKEKB collider at KEK in Japan~\cite{SuperKEKB} has a 3~km circumference. Future projects are of even larger scale, as considered for a linear collider \cite{CLIC, ILC}, or for circular colliders \cite{CEPC, FCCee}, where the ring circumference is expected to be larger than 90 kilometers. These accelerators require that diagnostics be properly synchronized. Accurate bunch identification, as required for the polarimetry of circulating beams, further necessitates precise timing \cite{FCCee, Charlet}. Very high stability has been achieved in accelerators for high-brilliance radiation sources at the European X-FEL \cite{SLAC, Kartner, XFEL, AllOptical, EOsync}. The distribution of time standards over more than 100~km has also been demonstrated in the past few decades \cite{NIST, NPL, LTE}. These costly all-optical systems or electro-optical systems are also implemented in existing colliders, where the most stringent requirement is imposed by the use of selective accelerating cavities~\cite{SKBsync, ATF}. A White-Rabbit protocol-based RF distribution is being considered as a reference for the high-luminosity upgrade of the LHC \cite{wlostowski:icalepcs2021-thbr02, Hazell}. Timing of large detectors with the accelerator also imposes some constraints on the stability of these links, but at a moderate level of picosecond precision \cite{LHCsync, LHCexp, Spring8}. The use of bunch-per-bunch beam position monitor also allows for the diagnosis with precision of dramatic events occurring in these large-scale colliders, detrimental to their performances, and where a picosecond timing system, easily integrable in an existing facility, is of interest \cite{BOR}. Furthermore, the use of polarized beams in future colliders and upgrades of existing colliders necessitates the implementation of laser-based diagnostics, such as Compton polarimeters \cite{EIC, ILC, CLIC, CEPC, FCCee, Charlet, BEPC2}. The use of modern laser technology, exploiting passively mode-locked laser oscillators with picosecond pulse durations, is considered for these projects. As a consequence, picosecond timing and synchronization of the laser with the few-picosecond-duration bunches of the accelerator is needed. The accelerator clock must be distributed over the scale of the accelerator, of several tens of kilometers for the largest. Existing systems employ the distribution of the absolute accelerator radio-frequency reference clock, the Master Oscillator (MO), through optical fibers, whose drifts are compensated for to achieve a precision of typically a hundred femtoseconds in the clock transfer, with latencies related to the fiber length \cite{SKBsync,Japan}. These systems are costly and also slightly overspecified for synchronizing the above-mentioned diagnostics. They are, moreover, difficult to implement, as they require some significant and expensive integration adjustments. The work described in this paper aims to demonstrate the proof of concept for a low-cost solution that achieves picosecond precision synchronization of a diagnostic integrated into an accelerator with minimal integration requirements. To that end, we utilize an electronics system developed at IJCLab, called \idro, that leverages the White Rabbit protocol~\cite{WrProject, WrPtpApp}. It is shortly described in section \ref{sec:idro}. This system is implemented in a laboratory room to synchronize a commercial laser system, representative of those used for accelerator diagnostics, but also for photocathode systems. The test setup is described in Section \ref{sec:setup}. Obtained results and prospects for improvements are further discussed in section \ref{sec:discussion}, before drawing conclusions.

\IEEEpubidadjcol
\section{The Idrogen system}\label{sec:idro}

\begin{figure}[b!]
    \centering
    \includegraphics[width=0.9\columnwidth]{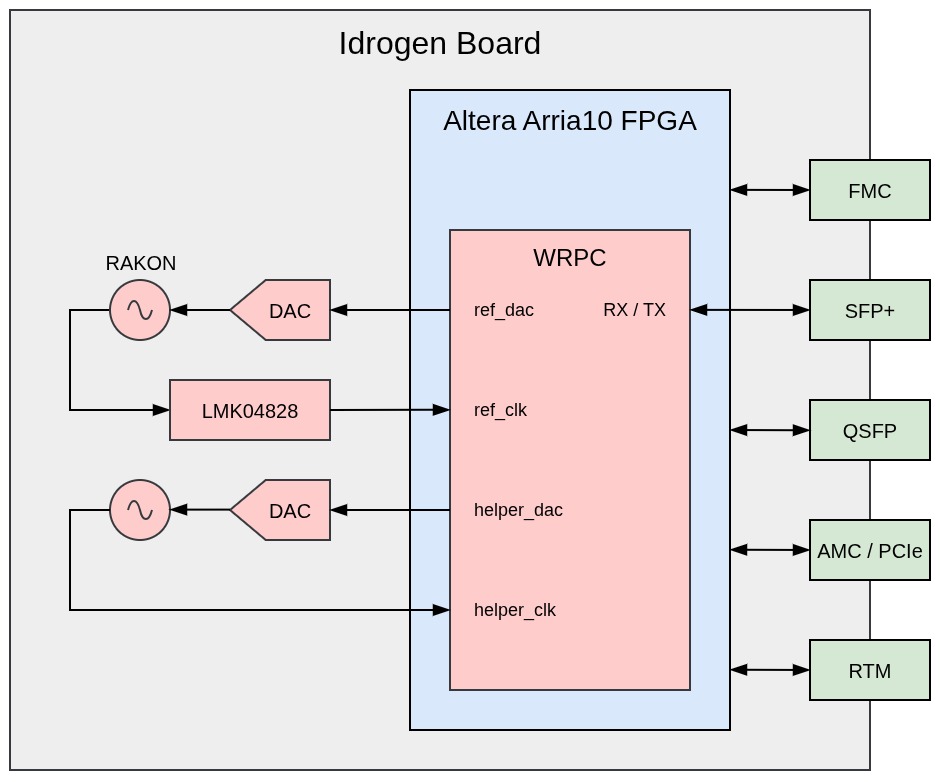}
    \includegraphics[width=0.9\columnwidth]{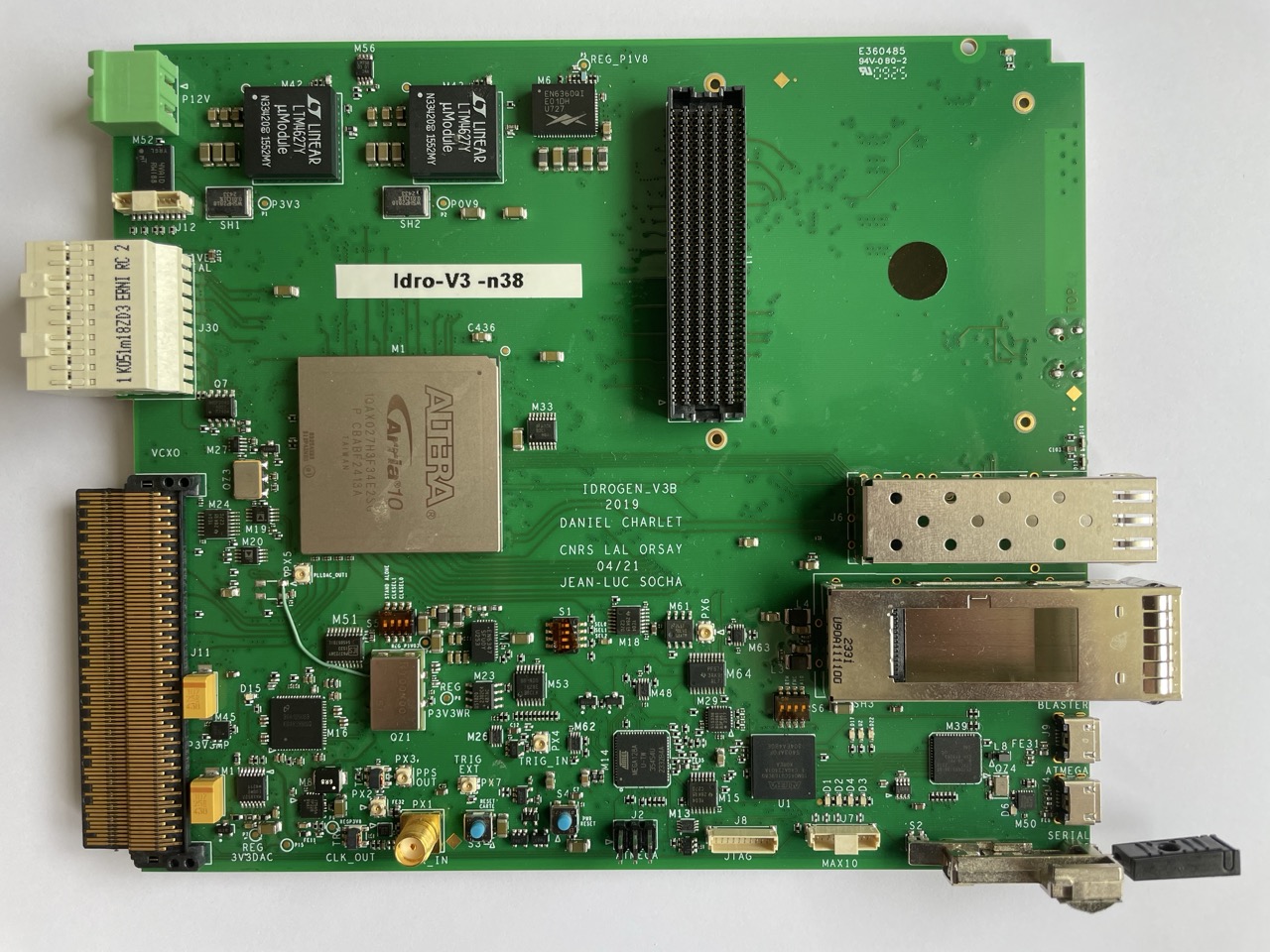}
    \caption{(top) Sketch of \idro\ board WR integration. (Bottom) Photograph of assembled \idro\ board. The outer dimensions are $15\times 10$~cm.
    }
    \label{fig:idrogen}
\end{figure}
\idro\ board is a \mtca\ board based on an Altera Arria\textregistered10 FPGA. This FPGA was chosen for its performance/price ratio. The main hardware elements of the \idro\ board are illustrated in Fig.~\ref{fig:idrogen}. At the heart of the \idro\ lies a high-speed FPGA that orchestrates all board functionalities, serving as the central controller for communication, timing, and signal processing tasks. This board has been designed to comply with the White Rabbit (WR) protocol~\cite{WrProject, WrPtpApp}.  
The White Rabbit protocol is an Ethernet based protocol designed to achieve high clock accuracy, stability, and resilience in a network. It extends the IEEE 1588-2008 Precision-Time-Protocol (PTP) [37], which aims at transmitting timestamp, and transfers via physical layer the reference frequency following the same concepts as the Synchronous Ethernet (SyncE) standard [36], i.e. Layer 1 syntonization. 
They rely on a Master-Slave network architecture. Synchronization is performed by a continuous delay measurement between master and slaves, using PTP, and a phase difference measurement, using a Digital Dual Mixer Time Difference (DDMTD)~\cite{ddmtd}, between a local oscillator and the clock recovered from physical layer. 
This continuous adaptation allows to compensate for variations in time propagation in the fiber due to the external environment, such as temperature. 
The White Rabbit network is therefore designed to distribute a sub-nanosecond synchronization (frequency and International Atomic Time, TAI) and data (Ethernet traffic) among a large number of nodes (more than 1000 nodes) across a large area (tens of kilometers). 
To improve phase sensitivity and simplify the FPGA design, the WR core (WRC) operates at 125~MHz. 
The WR core (WRC) is implemented within the FPGA of \idro\ board and can operate in both slave and (grand) master modes. 
The FPGA dynamically controls a low-noise 16-bit 20~kS/s digital-to-analog converter (DAC) to apply voltage tuning to the local 100~MHz oscillator, correcting frequency offsets according to the WRC.

Particular attention has been paid to provide high-stability performance by upgrading some components compared to the original implementation.
A critical feature of the design is an ultra–low–noise clock tree, built around the Texas Instruments LMK04828 jitter cleaner, which offers an outstanding root-mean-square jitter of 91 fs (integrated over 100 Hz to 20 MHz). 
The LMK04828 features an integrated VCO operating near 3 GHz and provides JESD204B-compliant clocking capabilities. 
Additionally, up to seven SYSREF signals are available to support high-precision acquisition synchronization for daughter cards or subsystems.

The LMK04828 is phase and frequency locked by a 100~MHz ultra–low-phase-noise oscillator (RAKON 1490U), featuring a phase noise of –116~dBc/Hz at 10~Hz offset — a specification we have carefully verified experimentally. 
The clock cleaner further accepts two external clock inputs, covering a frequency range from 10 MHz to 500 MHz, with a tunable loop bandwidth between 1 kHz and 100 kHz. 
This flexibility enables \idro\ to operate as a Grandmaster (GM) with exceptionally low additive phase noise when locked to an external reference.

The \idro\ board features one SFP+ port and one QSFP port. The SFP+ port is used to establish the 1Gb/s WR link. In parallel, the QSFP interface provides a data output capability of 40Gb/s, a critical feature for use cases requiring high-speed synchronous sampling, such as accelerator diagnostics, to name only one. 
A complete firmware and software ecosystem was developed to control and monitor the \idro\ board, including WRC parameters, using Ethernet or PCIe protocols. 

The \mtca\ implementation of \idro\ additionally provides full remote management and supervision capabilities, consistent with standard \mtca\ platform management frameworks. 
Furthermore, the generated clock signals can be broadcast over dedicated clock distribution channels reserved within the \mtca\ backplane, enabling precise timing dissemination across the crate.

The \idro\ board has been designed to be a generic and modular system. An FMC slot enables the extension of board capabilities, for example, with an ADC FMC board, allowing it to be used as a high data rate acquisition system synchronized by WR and compliant with JESD204B.
Additional functionalities can be integrated through dedicated RTM (Rear Transition Module) extension boards.
The implementation of the White Rabbit protocol within the FPGA enables precise time-stamping (4~ns) of all events acquired by the extension board, as well as of any signals or data streams processed internally by the FPGA.

\section{The test setup}\label{sec:setup}
\begin{figure*}[b!]
    \centering
    \includegraphics[width=0.9\textwidth]{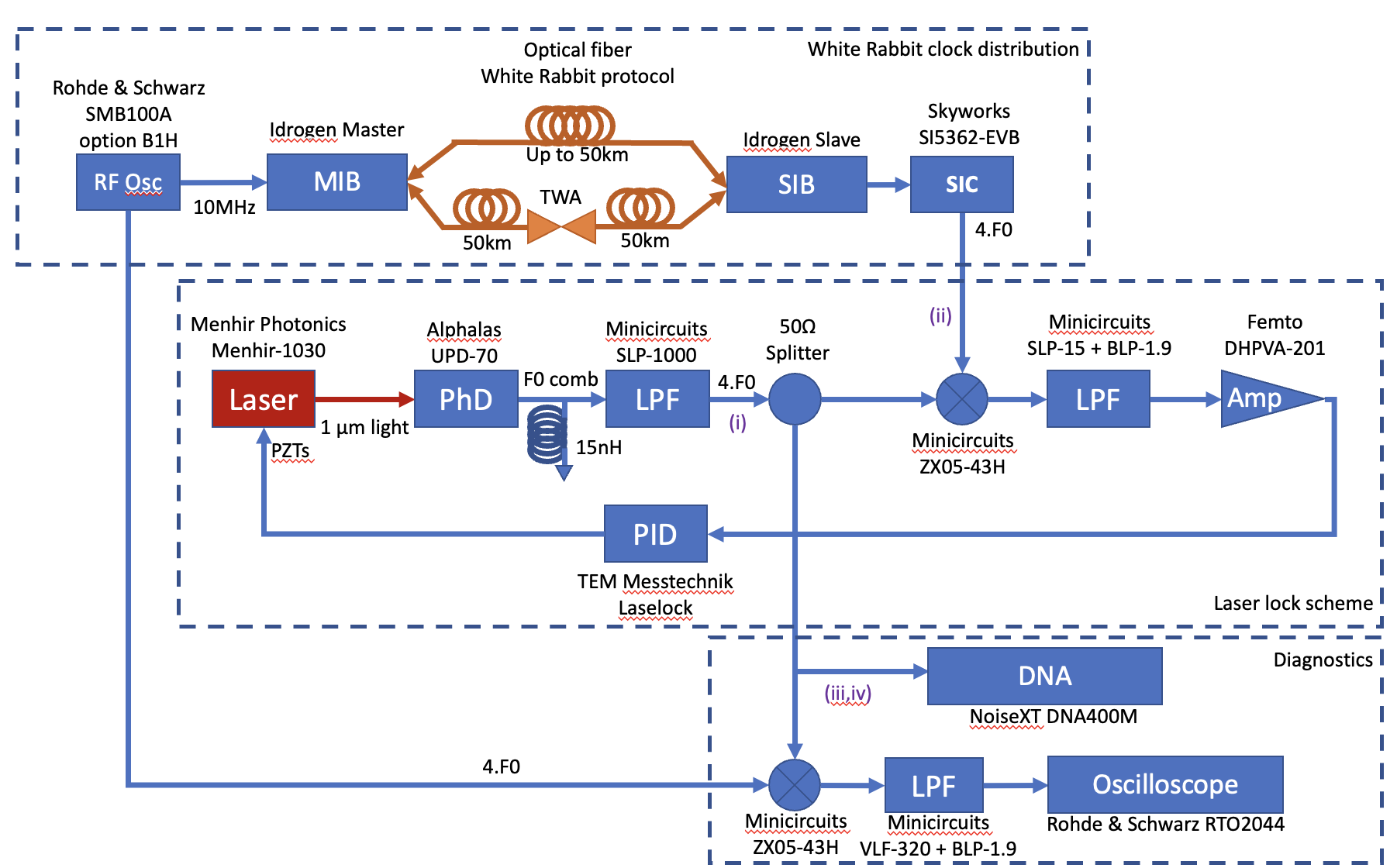}
    \caption{Diagram of the implemented test setup. Blue lines and elements are microwave components and links. Orange lines and components are fibered. Red line and element is the 1~µm laser. MIB: Master Idrogen Board. SIB: Slave Idrogen Board. SIC: Skyworks SI5362-EVB. TWA: Two-way optical amplifier. PhD: Photodiode to measure the laser light and produce a related microwave signal. PZTs: Piezo-electric transducers embedded in the laser, for its frequency adjustment. LPF: low pass filter. DNA: Digital noise analyzer, used to measure the Phase noise, PSDs and Allan deviations. It has been placed at different locations for the measurements reported in this paper. These are identified by the violet index (i,ii,iii,iv). The DNA clock reference, not shown in the figure, is either locked on the SMB signal generator for measurements (i, ii and iv) or the SIC output for measurement (iii). The two orange paths correspond to the two settings used: one for a fiber link of less than 50~km, the other for 100~km link.
    The 10MHz output of an SMB100A-B1H drives the clock of an Idrogen Master. Exploiting the White Rabbit protocol, the timestamp and reference clock are optically transferred to the slave Idrogen, where an RF signal at 250MHz is produced and fed to a SI5362-EVB, which embeds PLLs to generate a LVDS at 4$F_0$. It is mixed with the, split, filtered photo-detected light from the Menhir-1030 laser. It produces an error signal after low pass filtering and amplification, used in the PID corrector to drive the laser-embedded piezo transducers. The second split of the filtered photo-diode output is either measured directly with a phase noise analyser, or mixed with the 4$F_0$ output of the SMB100A-B1H and filtered to produce a measure of the phase difference between the measured RF signal from the laser and that of the reference signal generator. This phase detection is measured with an oscilloscope.}
    \label{fig:diagram}
\end{figure*}
This electronic system is installed in a roughly thermally stabilized room, typically used for laser experiments, where the temperature is expected to vary by a couple of degrees at most. The intention is to validate the performance of phase accuracy and stability by locking a MENHIR-1030 from MENHIR-PHOTONICS AG. This laser system is representative of passively mode-locked laser systems designed for use in Compton polarimeters in accelerators. It is
equipped with two piezo-electronic transducers (PZTs) for the tuning of the repetition rate around its nominal value of $F_0=216.661555~MHz$. The diagram corresponding to the implemented setup is given in Fig.~\ref{fig:diagram}. To validate the performance of the synchronization system, the master \idro\ board (MIB) is disciplined by a 10 MHz reference from a Rhode \& Schwarz SMB100A microwave signal generator equipped with an OCXO B1H. Another \idro\ board (SIB), locked through a White Rabbit link to the MIB, generates a 250 MHz LVDS signal that is provided to a SI5362-EVB-A from SkyWorks (SIC). The fiber link between MIB and SIB can be configured to be either with two 10~m, 5~km, 50~km fibers or a single 100~km long fiber. In the 100 km configuration, the total optical loss budget is significantly affected by fiber attenuation as well as connector and splice losses accumulated across the experimental infrastructure, resulting in a received optical power approaching the sensitivity limit of the transceivers.
To ensure reliable bidirectional communication under these conditions, a bidirectional optical amplifier is inserted in the middle of the link. This configuration ensures sufficient optical margin for stable operation of the synchronization system and is representative of realistic long-distance deployments, where heterogeneous fiber infrastructures and intermediate amplification stages are commonly encountered in field implementations of White Rabbit–based networks. The SI5362 chip embedded on the SIC allows arbitrary frequency generation, with Hertz precision, synchronous to the SIB output. The SIC is configured with two outputs that deliver clocks at the fourth harmonic, $4F_0$, of the laser frequency and one at $F_0$. One of the $4F_0$ outputs is passively mixed (ZX05-43H from Minicircuits) with the microwave signal obtained by sending the laser light to a UPD-70 photo-diode from Alphalas GmbH, amplified, and filtered. The choice of filters is not optimal, but it reflects the limited component selection available at the time of measurement. A quick check of the measured spectrum with a spectrum analyzer shows that the $4F_0$ is the largest contribution, with extinction of other harmonics by at least 20~dB. The mixed output is filtered and used as an error signal to seed a commercial PID corrector. The LaseLock$^{\mbox{\scriptsize{\textregistered}}}$ from TEM Messtechnik GmbH is configured to drive both PZTs of the laser. A split of the amplified output of the laser photo-diode is used for diagnostics, such as long-term phase drift and short-term phase noise power spectral density measurement. 

\begin{figure}[t]
    \centering
    \includegraphics[width=\columnwidth]{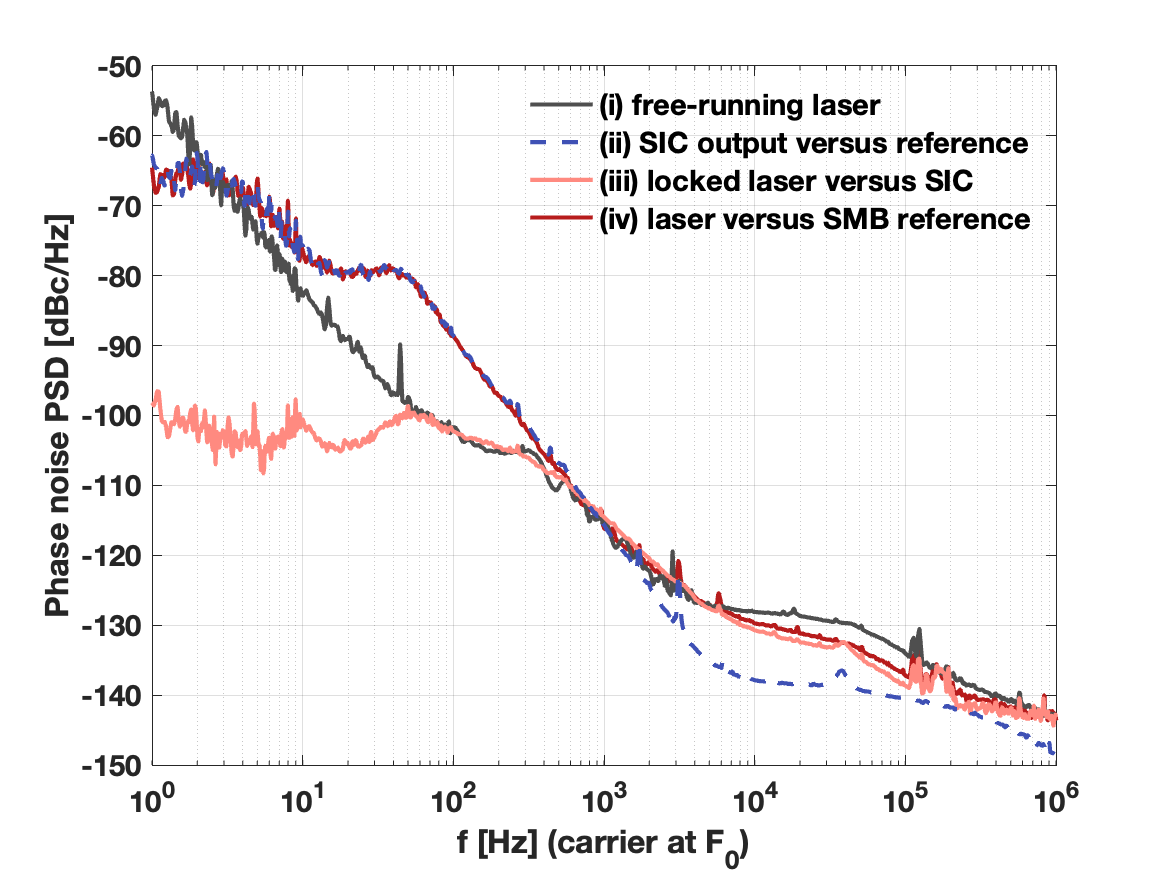}
    \caption{Power spectral density of phase noise for the (i) free running laser; (ii) the SIC output with respect to the reference; (iii) the technical noise of the laser loop on the SIC output; (iv) the laser output with respect to the reference. A 10~m long fiber link was used in between the Idrogen boards for these measurements.
    }
    \label{fig:PNPSD}
\end{figure}

\begin{figure}[t]
    \centering
    \includegraphics[width=\columnwidth]{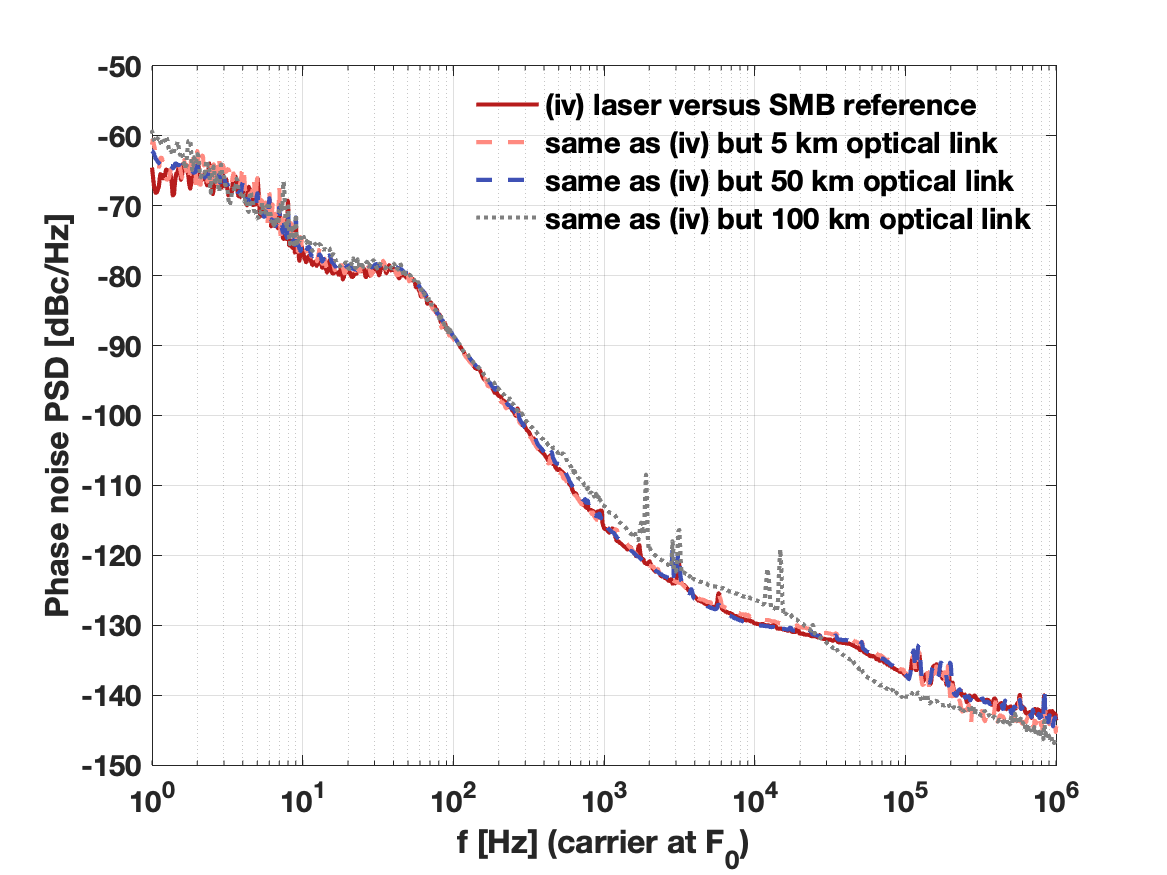}
    \caption{Power spectral density of phase noise for the (iv) the laser output with respect to the reference with 10~m, 5~km, 50~km and 100~km fiber links. A difference can be spotted with the 100~km where the noise is slightly larger above 300~Hz. Preliminary investigations suggest that this might be due to a change of the measurement setup where amplifiers have been added after the 50 Ohm splitter.
    }
    \label{fig:longfiber}
\end{figure}

\begin{figure}[htbp]
    \centering
    \includegraphics[width=\columnwidth]{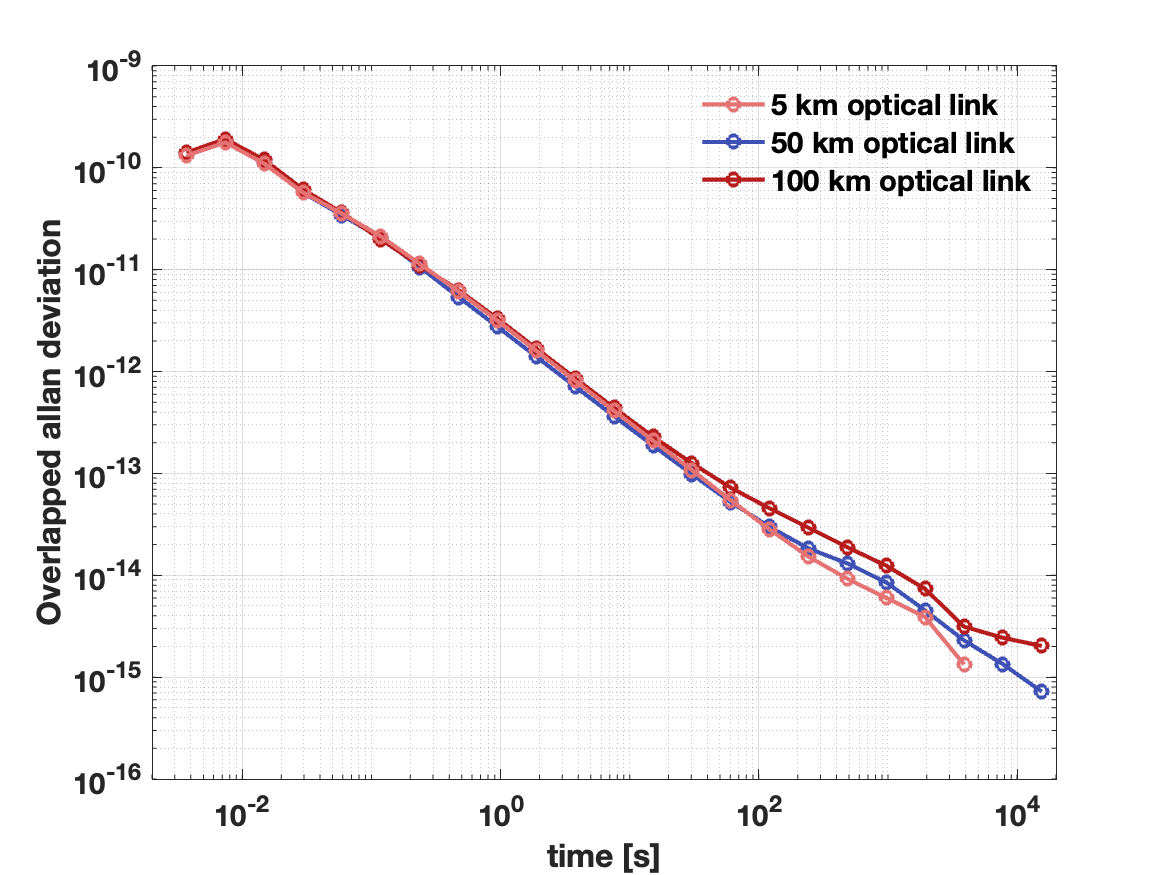}
    \caption{Measurement of the overlapped allan deviation of the laser locked on the SIC with respect to the SMB reference, for various fiber length. 
    }
    \label{fig:allan}
\end{figure}

The power spectral density of phase noise is measured with a NoiseXT DNA-400M Digital Noise Analyzer (DNA). Its bandwidth is limited to 400~MHz, so only microwave signals at $F_0$ are measured. Several measurements are performed: (i) the free-running laser output without the amplifier after the photodiode; (ii) the output of the SIC at $F_0$; (iii) the output of the laser locked on the SIC, measured with respect to the SIC reference at $F_0$; (iv) the output of the laser, locked on the SIC, measured with respect to the SMB reference at $F_0$. The results are shown in Fig.~\ref{fig:PNPSD}. The computed integrated phase noise for (iv) is 1.4 ps RMS over the measurement range. We note the excellent consistency of the measurement of the locked laser onto the SIC, the SIC output, and the added technical noise by the laser/SIC loop. We have checked that the phase noise of the free-running laser measured after the splitter follows the curve (iii) above 50~Hz. It results in an apparent degradation of the exquisite laser phase noise due to the measurement chain rather than a genuine degradation of the laser phase noise. A measurement of the laser phase noise in the optical domain is necessary to conclude on this aspect, which is out of the scope of this proof of concept paper. These results state the current performance of the system. It has been reproduced with an E5052A SSA from Agilent. These results were obtained with a 10~m long fiber link between the MIB and SIB boards. As can be seen in Fig.~\ref{fig:longfiber}, we observe no modification of the power spectral density of residual phase noise when replacing it with a 5~km and 50~km fiber link because the deviation frequency range is limited to 1~Hz at low frequency. However, the use of a 100~km fiber link exhibits two differences. At low deviation frequencies, around 1~Hz, the system seems slightly less stable. It might be due to the addition of an optical amplifier at the midpoint of the fiber to allow enough signal to reach the transceivers. Above 300~Hz a modification of the phase noise spectral density is observed. This confirms the assumption that this part of the spectrum is rather dominated by technical noise, as amplifiers were added on both outputs of the 50 ohm splitter. The overlapped Allan deviation is measured to be approximately $3.1\times10^{-12}$ at 1~s with a regular $\tau^{-1}$ decay for the laser synchronized to the SIC, measured relative to the SMB reference, as shown in Fig.~\ref{fig:allan}. A departure of the Allan deviation on timescales of about 10~s is observed for the 100~km fiber link (including the fiber amplifier), compared to other measurements. This effect is left for further investigations.

\begin{figure}[t]
    \centering
    \includegraphics[width=\columnwidth]{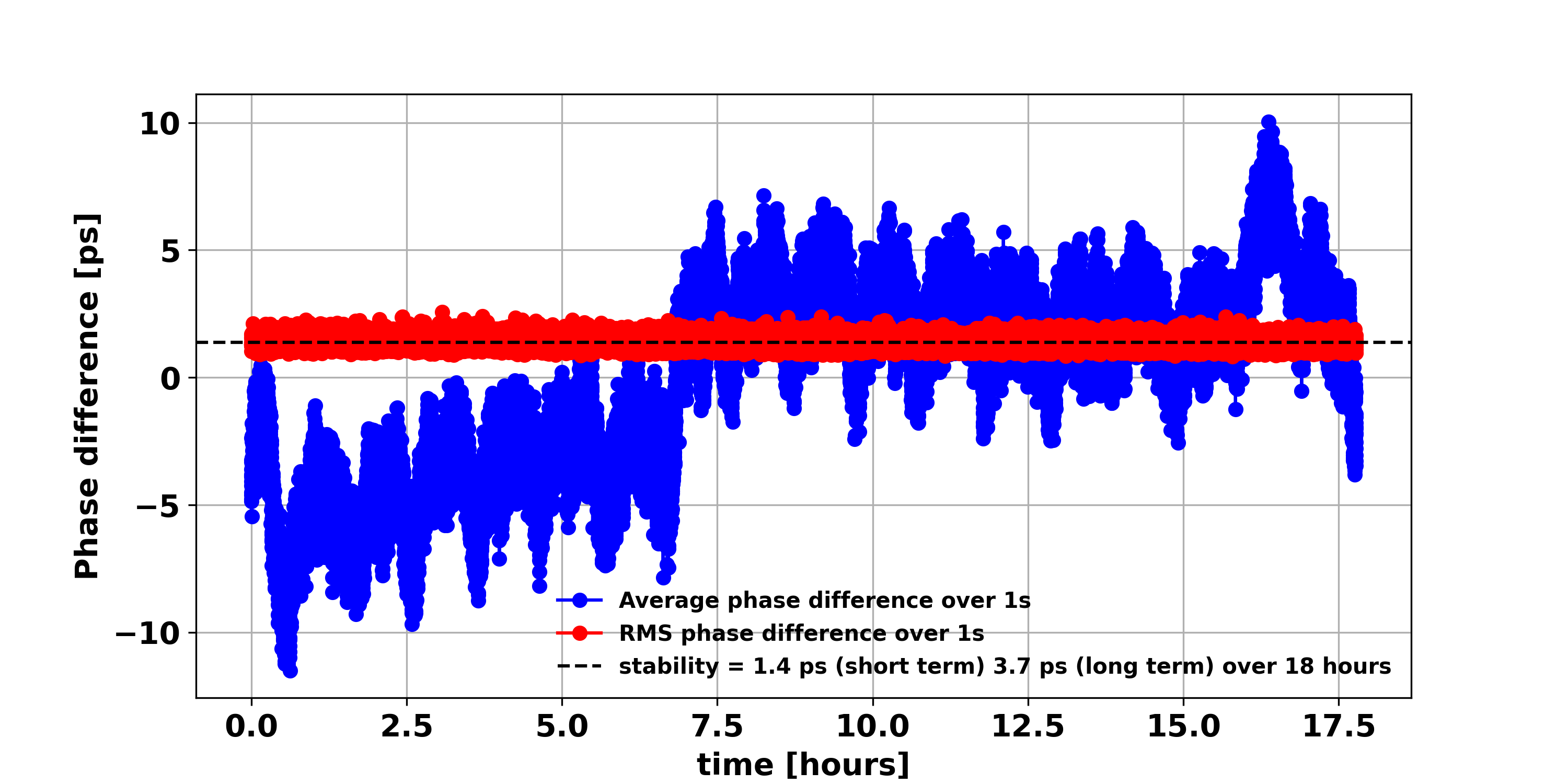}
    \includegraphics[width=\columnwidth]{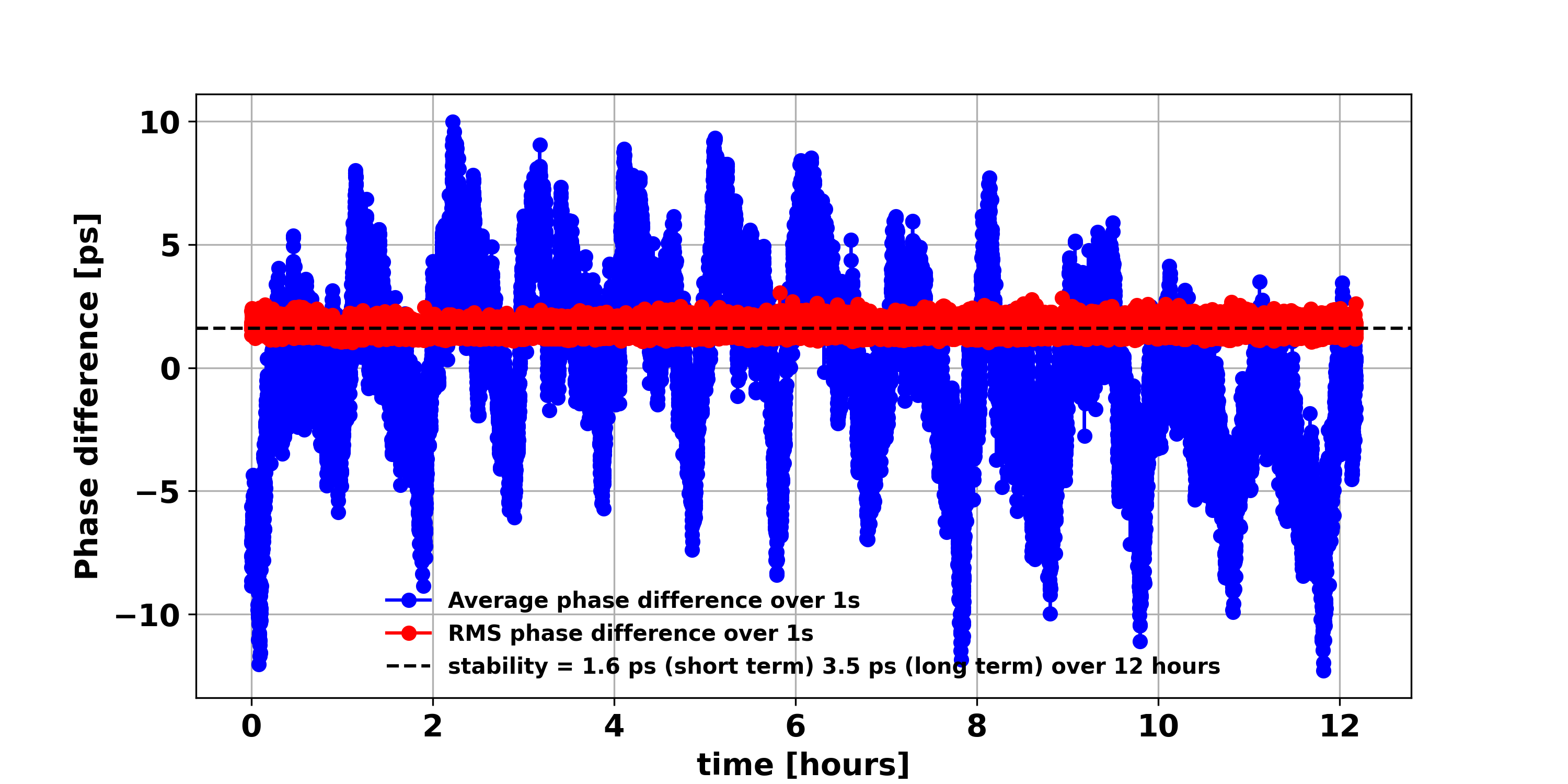}
    \includegraphics[width=\columnwidth]{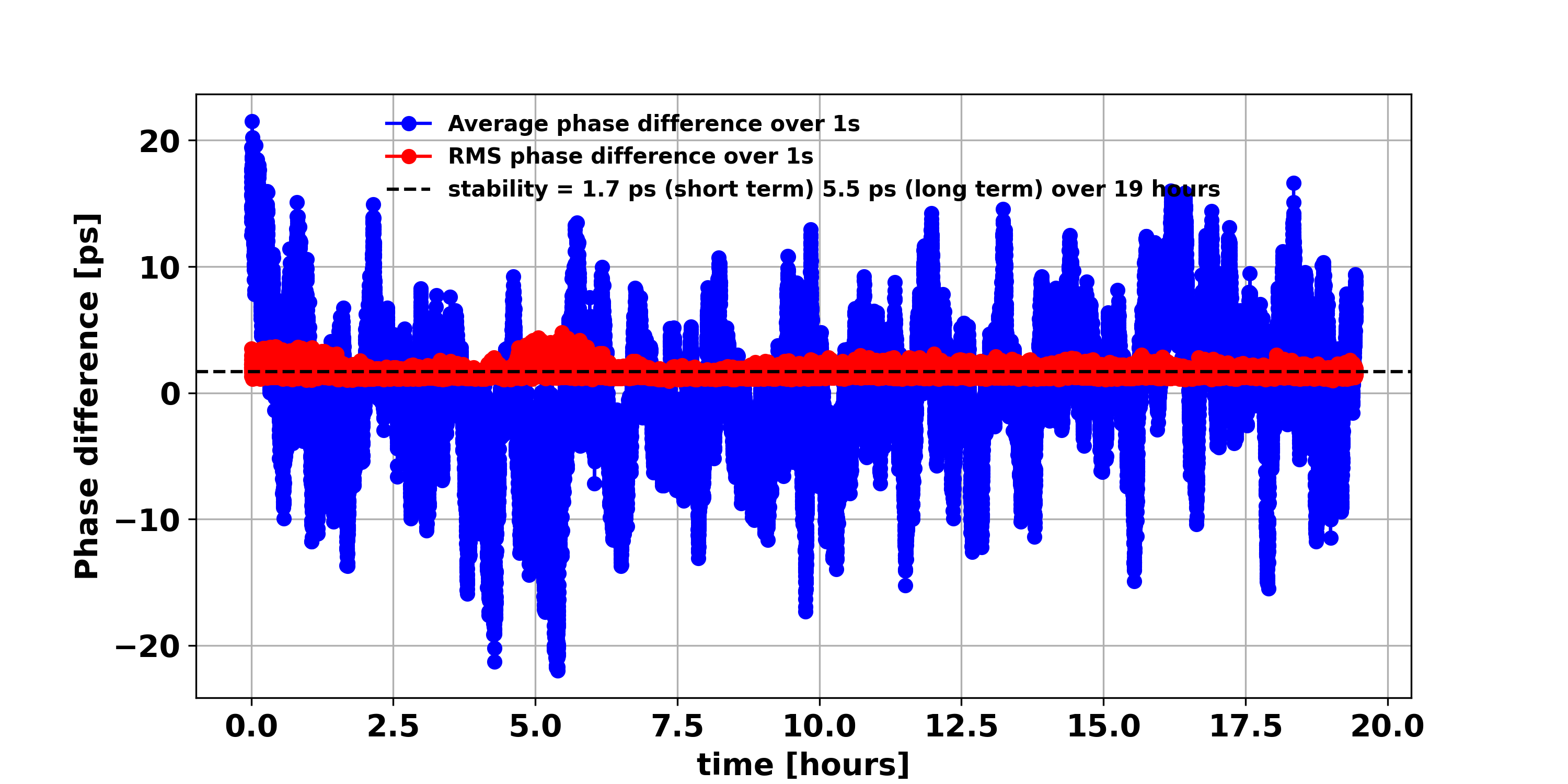}
    \caption{The measurements over 17, 12 and 19 hours of the mean and RMS phase difference (converted in picoseconds) measured over one second (forty million samples) between the laser locked on the SIC and the SMB reference, virtually located at 5 (top), 50 (middle) and 100 (bottom) kilometers away from the laser.
    }
    \label{fig:phase_data}
\end{figure}

\begin{figure}[t]
    \centering
    \includegraphics[width=\columnwidth]{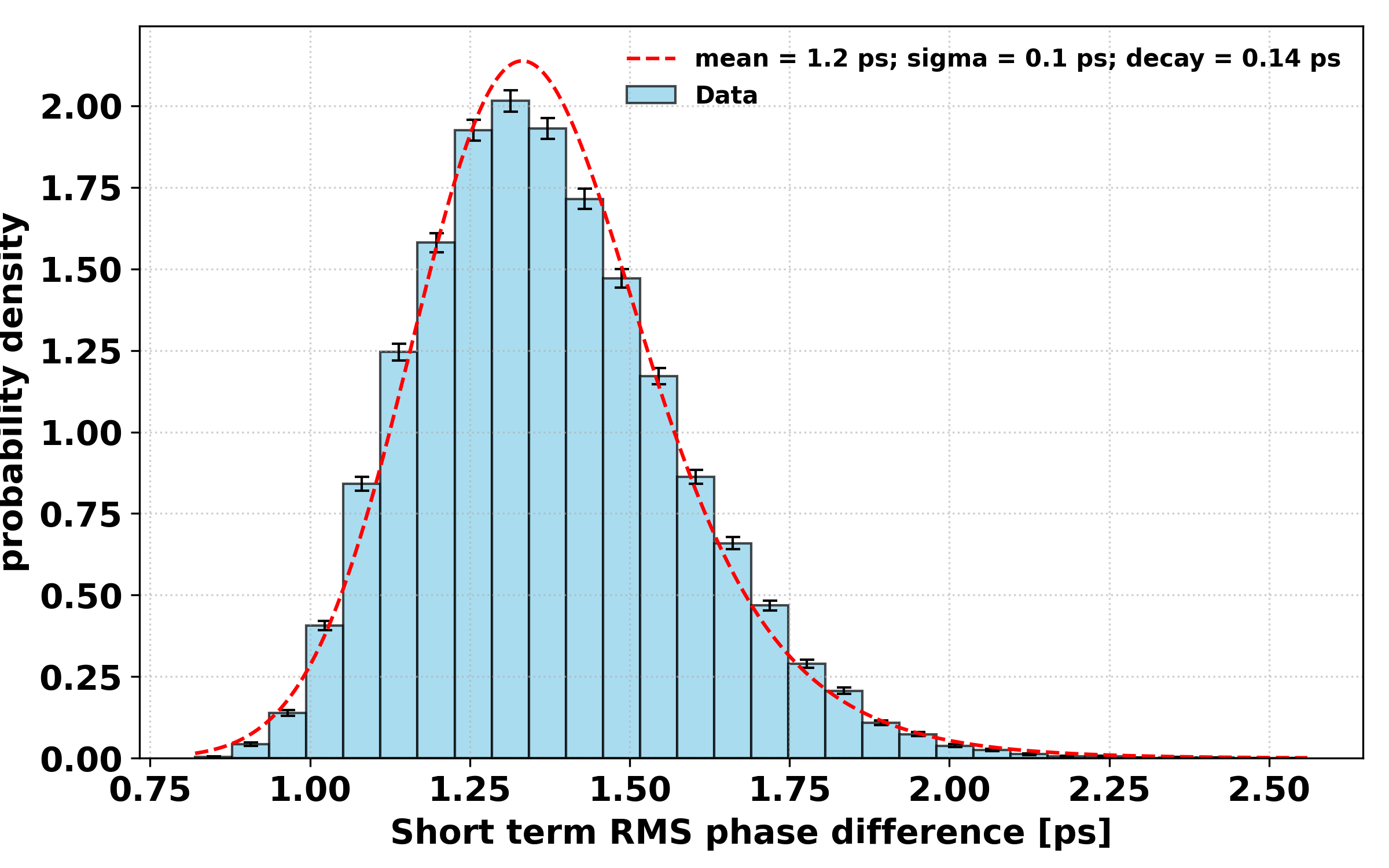}
    \includegraphics[width=\columnwidth]{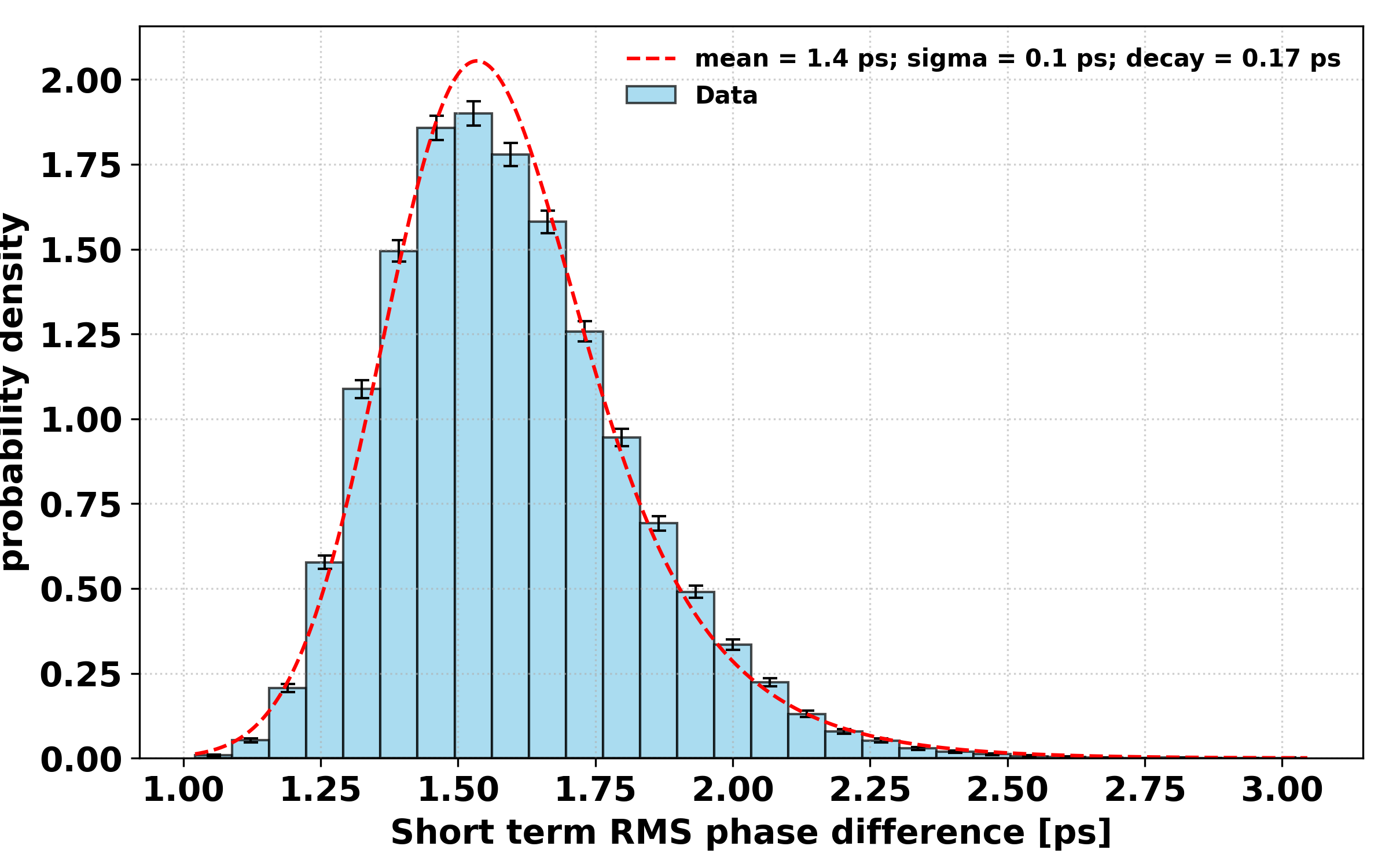}
    \includegraphics[width=\columnwidth]{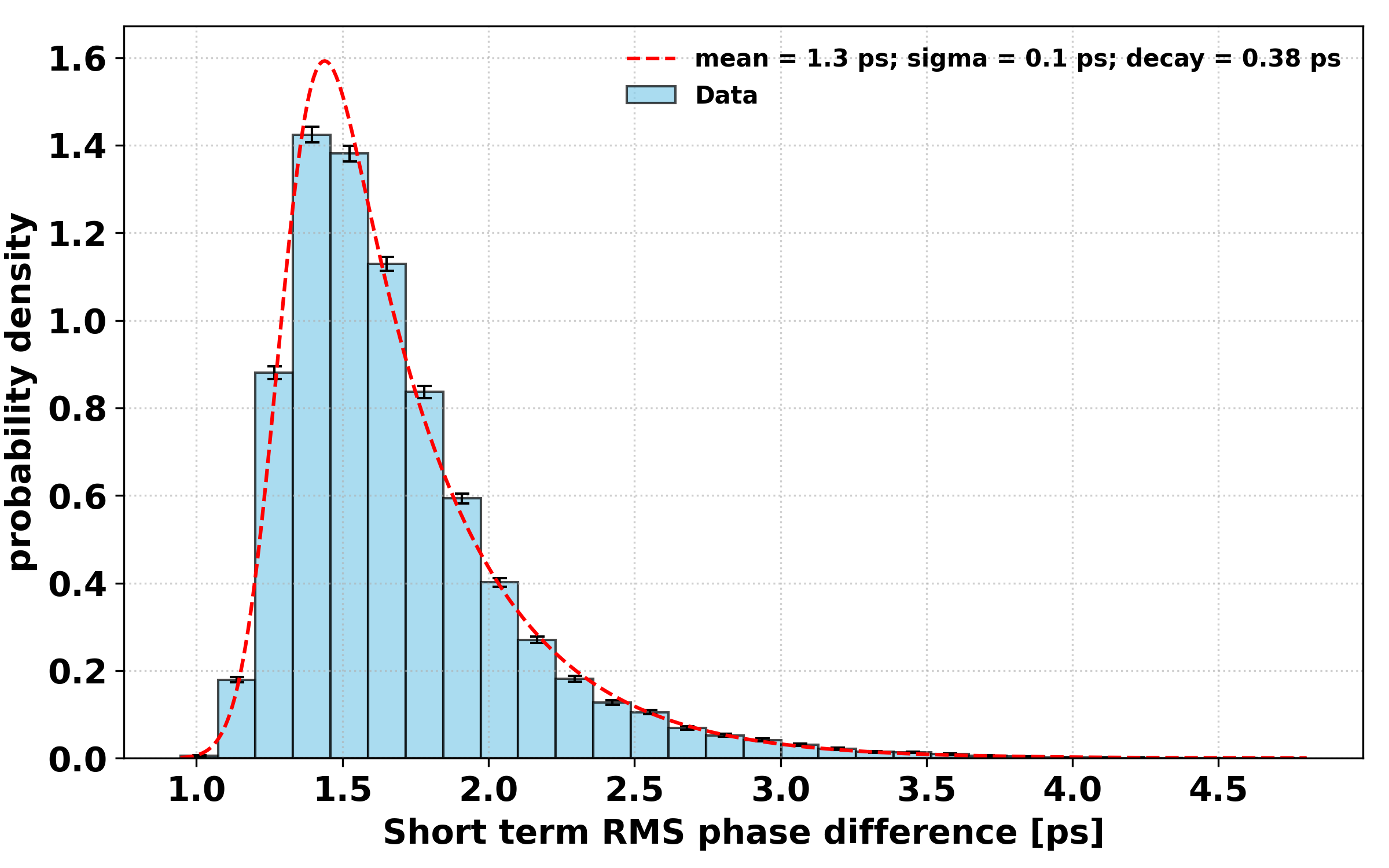}
    \caption{Histogram of the measured RMS phase difference (converted in picoseconds) in blue stacks with black error bars, for 5~km (top), 50~km (middle) and 100~km (bottom), respectively. An exponentially convoluted Gaussian distribution (read dashed line) is fit to the data. It provides a mean of 1.2~ps (5~km), 1.4~ps (50~km) or 1.3~ps (100~km), a width of 0.1~ps and a decay constant of 0.14~ps (5~km), 0.17~ps (50~km) or 0.38~ps (100~km).
    }
    \label{fig:phase_hist}
\end{figure}

To further complement these measurements, a phase detection is realized between the amplified RF signal from the photodiode and the SIC 866.66 MHz output using a passive mixer and filter. It has been realized for the various fiber lengths. Results with 5~km, 50~km and 100~km are presented in the following. The phase is acquired with an oscilloscope. While opening the laser/SIC locking loop, a clean sine signal is observed with a large amplitude of about 700~mV. We then close the loop and measure the calibrated phase difference. Waveforms of one second duration are acquired and analyzed in real-time by the apparatus to extract the standard deviation over a second, and the average phase also over a second. It is tracked for several hours and shown in Fig.~\ref{fig:phase_data}. We observe excellent reproducibility of the standard deviation of the phase, which is consistent with the measured phase noise of the DNA, as shown in Fig.~\ref{fig:phase_hist}. The pulse-per-second (PPS) also provides a consistent picture. We compute the short-term stability as the average over a large number of measurements of the standard deviations of phase over 1~s. The long-term stability is computed as the standard deviation of the average phase over 1~s. 
For completeness, the Time Deviation (TDEV) has been derived from the dataset presented in Fig. 6, using a sampling interval of 2~s. The TDEV provides a complementary view of the timing stability, particularly at longer integration times where noise processes associated with the fiber link become more apparent.
The measured TDEV reaches 0.7 to 0.9 ps at 2 s integration time and 0.9 to 3 ps at $10^3$ s integration time, depending on the fibre length. These results are consistent with the stability inferred from the overlapping Allan deviation and confirm that the system maintains picosecond-level timing stability over the 100 km link. 
Variations with a period of approximately one hour, likely related to the cycling of the room's air conditioning. As a consequence, the long-term, stability is significantly and consistently larger than the short-term stability.

\begin{figure}[htbp]
    \centering
    \includegraphics[width=\columnwidth]{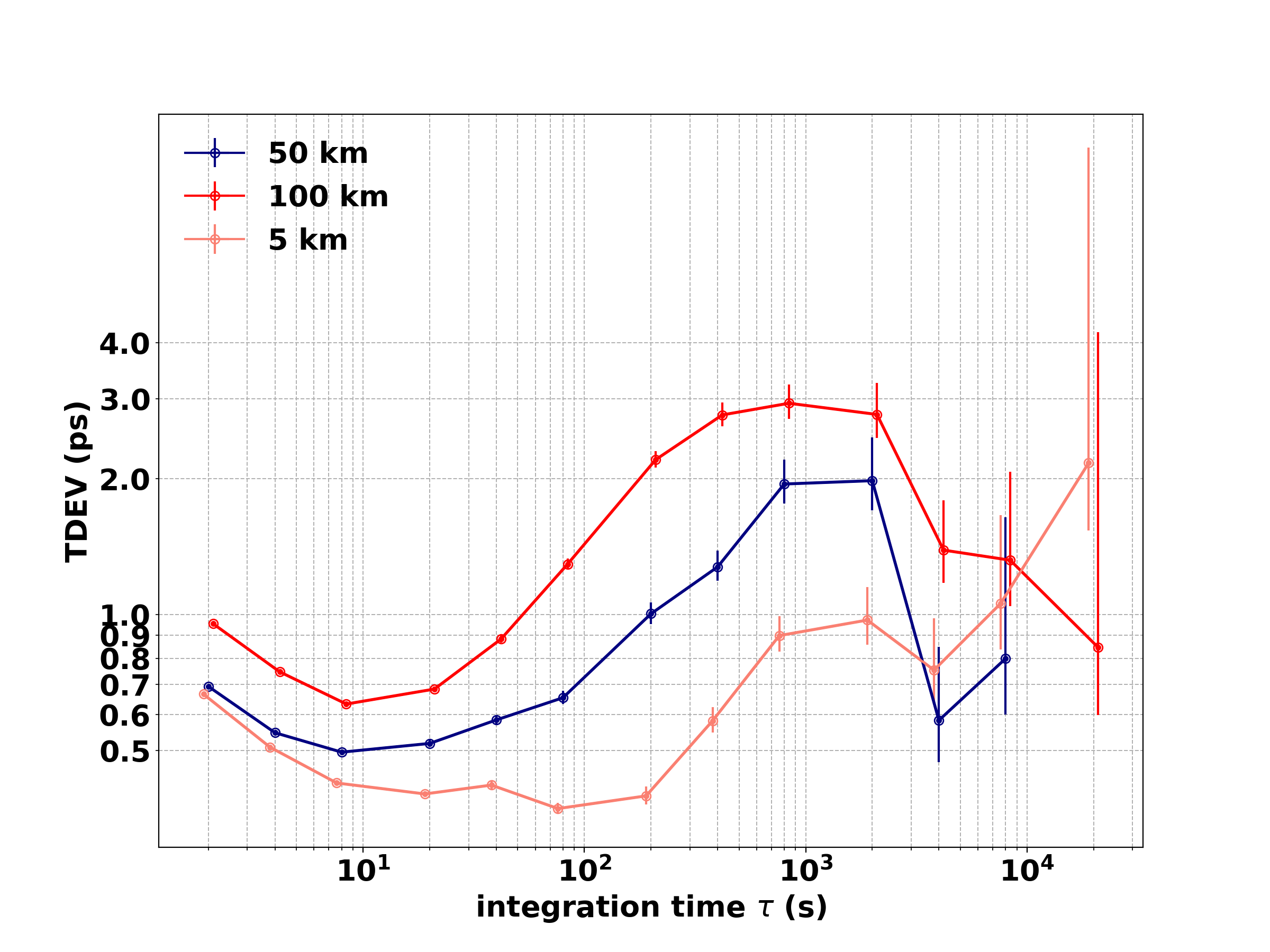}
    \caption{Measurement of the time deviation of the laser locked on the SIC with respect to the SMB reference, for 5, 50, and 100~km fibre lengths. Data points for 5 and 100~km curves are slightly horizontally shifted to better visualize the points and error bars.
    }
    \label{fig:tdev}
\end{figure}

These likely explain the observed variations. Note that during these measurements, the boards were not encapsulated in housing with thermal control, nor in \mtca\ crates. The results with the 100~km were found to exhibit long-term stability worse by a factor larger than 1.5 compared to 50~km data, at 5.5 ps RMS. The TDEV is also getting worse when increasing the fibre length, which also suggests some residual uncompensated drift related to the fibre. The distribution of the measured RMS phase differences gets more peaky at 1.4~ps with a longer tail at larger RMS values, suggesting a less stable operation at 100~km. An exponentially convoluted Gaussian distribution \cite{Grushka} is found to fit the data well and is motivated by the observation that thermal variations induce a shift in the phase difference with some finite relaxation time. It allows for representing the presence of a positive definite additive delay. The obtained mean of the Gaussian is consistent with the RMS jitter computed from the phase noise PSD, as expected. The time over which the laser could be kept locked was found very different from one day to another, likely due to different environmental conditions. Further investigations, with encapsulated or racked boards, are kept out of the scope of this proof of concept paper, and will allow to conclude on this aspect.

\section{Discussion}\label{sec:discussion}
The obtained results demonstrate that picosecond-level short-term stability (over 1~s) has been achieved for two microwave clocks, virtually separated by one hundred kilometers of regular fiber link, using the Idrogen electronic system, which uses the White Rabbit protocol. The long-term stability over half a day is found to be worse, up to 5.5~ps for the 100~km link. Accuracies of the phase difference corresponding $\pm 10$~ps and $\pm 20$~ps are obtained for the 50~km and 100~km fiber links, respectively. Consistency is achieved between the measurements of the power spectral density of phase noise and phase difference with respect to the reference clock at 866 MHz. The apparent degradation of laser phase noise when in the measurement chain, above 50~Hz, might be solved in the future for some dedicated applications that require exquisite phase noise \cite{ThomX, GFpop}. 
The test was conducted with boards simply placed on a table, and neither encapsulated in housing nor integrated into racks, as would be necessary for integration in an accelerator. Similarly, fibers were simply placed in the room, without additional temperature control with respect to the room air conditioning. Further tests with temperature control of the system elements are out of the scope of this paper. The observed drift of the relative phase over hours of operation is likely due to environmental variations. These issues will likely be mitigated with proper housing and venting of the boards or incorporation into \mtca\ crates, which remains to be tested but is outside the scope of this proof-of-principle work. Preliminary investigations suggest that this effect is mainly related to the SI5362-EVB. Implementation of corrections for known temperature changes, exploiting probes embedded on the \idro\ boards, is planned. The work also demonstrates that the lock of the laser was preserved for 16 hours without loss, a duration limited here by the allocated time for the experiments and the availability of the hardware. Longer tests will be performed at a later stage. The development of an FMC with the SI5362 chip is also underway to fully leverage the modularity of the \idro\ system, resulting in a compact, low-cost, and arbitrary frequency generator with sub-picosecond timing that is easily integrated into an accelerator environment. The hope is that these developments will significantly help understanding and possibly improve the long term stability of the system.

\section{Conclusion}
The results presented in this paper demonstrate that a frequency generation method that is based on the \WR\ protocol, integrated on the \idro\ board with ultra–low–noise components, enables achieving picosecond precision over a 1-s window. The long-term stability over half a day is found to be 5.5~ps over a 100~km distance. The accuracy of the phase difference corresponding to $\pm 20$~ps is obtained. The present result already leverage the possibility to implement this system in accelerator for further experimental performance validation, as synchronization of photocathode laser systems for beam generation~\cite{ATF}, Compton polarimeters \cite{Charlet} and interaction with beams \cite{GFpop, ThomX}, and acquisition of bunch-per-bunch beam profile monitors, exploiting the capability of \idro\ to embed a 1~GHz FMC ADC \cite{BOR}. It further paves the way for the use of this technology for frequency generation with a few picoseconds stability in large-scale facilities, where scalability is a crucial aspect \cite{scalability}. Large detectors located on large colliders also represent an area of application of the concept evaluated here \cite{TimingDet2}, with possible implications in cosmic ray detection and medical physics \cite{TimingDet}. 

\section*{Acknowledgments}
The authors thank K. Popov from KEK ATF, iCASA, and C. Joly from IJCLab for fruitful discussions. Part of the instrumentation
was funded by the “Investissements d’Avenir” programme launched by the French Government and implemented by the Agence Nationale de la Recherche (ANR) under reference
ANR-21-ESRE-0029 (ESR/Equipex+ T-REFIMEVE). This research has been funded, in whole or in part, by  l'Agence Nationale de la Recherche (ANR), project ANR-25-CE31-7501. For the purpose of open access, the author has applied a CC-BY-NC-ND public copyright
licence to any Author Accepted Manuscript (AAM) version arising from this submission.

 

%



\end{document}